\documentclass[twocolumn,preprintnumbers,amsmath,amssymb]{revtex4}

\usepackage{graphicx}
\usepackage{dcolumn}
\usepackage{bm}

\begin{document}


\title{Non-modal approach to linear theory: \\marginal stability and the dissipation of turbulent fluctuations}

\author{Enrico Camporeale}
\altaffiliation{Queen Mary University of London, Mile End Road, London E1 4NS, United Kingdom}

\author{Thierry Passot}
\altaffiliation{University of Nice Sophia Antipolis, CNRS, Observatoire de la C\^{o}te d'Azur, B.P. 4229, 06304 Nice Cedex 4, France}

\author{David Burgess}
\altaffiliation{Queen Mary University of London, Mile End Road, London E1 4NS, United Kingdom}

\date{\today}

\begin{abstract}
The non-modal approach for a linearized system differs from a normal mode analysis by following the temporal evolution of some perturbed equilibria, and therefore includes transient effects. We employ a non-modal approach for studying the stability of a bi-Maxwellian magnetized plasma using the Landau fluid model, which we briefly describe. We show that bi-Maxwellian stable equilibria can support transient growth of some physical quantities, and we study how these transients behave when an equilibrium approaches its marginally stable condition. This is relevant to anisotropic plasma, that are often observed in the solar wind with a temperature anisotropy close to values that can trigger a kinetic instability. The results obtained with a non-modal approach are relevant to a re-examination of the concept of linear marginal stability. Moreover, we discuss the topic of the dissipation of turbulent fluctuations, suggesting that the non-modal approach should be included in future studies.\\
\\
\end{abstract}

\maketitle

%
%

%

%
%

\section{Introduction}
Linear theory represents a powerful tool for the interpretation and understanding of many space plasma properties observed by in situ spacecraft. For instance, the temperature anisotropy of the solar wind is thought to be bounded by values that are consistent with the stability thresholds derived within the linear theory of the Vlasov-Maxwell set of equations. The broadly accepted view is that a macroscopic property of the plasma (such as temperature, density, or mean velocity) can be constrained by the nonlinear feedback associated with a linear instability. This is because the primary consequence of any instability is to reduce the amount of free energy that drives the instability, relaxing the plasma towards a marginally stable condition. This is equivalent to saying that the plasma is unlikely to be found in an unstable state because it tends to change its macroscopic properties in a way that would lead to a linearly stable condition.\\
In the solar wind it is argued that the expansion of the plasma from the Sun in a radially decreasing magnetic field should produce much higher values of temperature anisotropy (with respect to the background magnetic field) than those observed. Also, it has been shown that the highest values of anisotropy observed in the solar wind are consistent with the thresholds of linear kinetic instabilities driven by temperature anisotropies. This is true both for protons \citep{kasper02, hellinger06, matteini07} and electrons \citep{gary05,stverak08}, for a large range of plasma beta, and both for $T_\perp/T_\parallel>1$ (whistler and mirror instabilities), and for $T_\perp/T_\parallel<1$ (firehose instability). \\
The physical behaviour of an unstable anisotropic collisionless plasma subject to the electron firehose instability, that relaxes towards marginal stability, has been elucidated in \citet{camporeale08}, with fully non-linear Particle-in-Cell simulations. What emerges is that the plasma state is likely to be found bouncing around the marginal stability threshold, due to the competition of two mechanisms: the reduction of the anisotropy and plasma beta caused by the development of the firehose instability (above the threshold), and the increase of these quantities due to the damping of magnetic fluctuations (below the threshold) that result in the energization of the particles, predominantly in one direction.\\
Despite the success of the linear theory to delineate the macroscopic properties in which the solar wind plasma is more likely to be found (i.e. in stable conditions), and the good agreement between linear theory predictions and solar wind data, at least two contradictions remain unquestioned.\\
First, the greater part of the protons and electrons in the solar wind, at any distance from the Sun, is observed to be isotropic or very lightly anisotropic \citep{hellinger06,stverak08}, i.e. in a condition where no anisotropy instability can be excited and therefore the aforementioned argument associated with the linear constraining mechanism cannot be invoked. In other words, the fact that linear instabilities do not allow the plasma to develop anisotropies higher than certain values, does not explain why most of the plasma is found to be very far from those values, since the effect of the expansion should result in a continuous increase of the anisotropy. Observational results have suggested, as a possible explanation, that the collisional age is related to the isotropization of thermal electrons \citep{salem03,stverak08}, but the relative importance of collisions and instabilities is still unknown.\\
Second, a relatively high occurrence of short wavelength magnetic fluctuations with small amplitude is persistently found in the plasma, even in stable conditions. Unless one assumes that a turbulent cascade is able to produce such fluctuations at a rate that perfectly balances the dissipation, this might be in contradiction with the notion that a linear perturbation in a stable plasma would damp exponentially in time.\\
The aim of this paper is to present a non-modal approach to the linear stability problem for a collisionless plasma in an uniform magnetic field that, by offering a new framework for the understanding of linear marginal stability, will reconcile those two apparent observational contradictions with a consistent physical interpretation.\\
Another important application of the plasma linear theory that will be revisited in the light of a non-modal approach is the interpretation of the physical mechanism that controls the kinetic damping of fluctuations in the turbulent dissipation range. This is an area of increasingly active research, and it has been argued that the linear approximation (i.e. the assumption that perturbed quantities have a much smaller amplitude than the equilibrium ones) might be used in this scenario \citep{gary04a}. There is a general agreement on the fact that the nonlinear cascade of turbulent fluctuations should take into account the onset of kinetic effects at a certain spatial scale, around the ion Larmor radius \citep{matthaeus08}. It is indeed observed that the power density of magnetic fluctuations undergoes an abrupt steepening for wavenumbers above a certain value \citep{leamon98a,hamilton08,sahraoui09}, and it is thought that kinetic effects might be responsible for the steepening. However, a complete understanding on exactly why this happens and what are the important parameters that determine at which scale one should expect the kinetic effects to produce a steeper slope of the power spectrum (which is observed to follow a power law in wavenumber both above and below the steepening) is still missing.\\
Altough the study of kinetic instabilities, and their application to the solar wind plasma, has been traditionally kept separate from the studies addressing the turbulent dissipation range, we look at those two aspects of plasma physics as interlocked. In fact, they describe the same problem from two distinct viewpoint, because the understanding of turbulent dissipation at small scales can be properly addressed only by including in the energy balance also the injection of turbulent fluctuations due to the development of kinetic instabilities.\\
Different approaches to this problem include the use of gyrokinetic linear theory \citep{howes06,howes08a} and simulations \citep{howes08b}, Hall-MHD \citep{krishan04,galtier07,shaikh09} and Particle-in-Cell simulations \citep{gary08}. A simple diffusive model for the nonlinear cascade in the inertial range (i.e. for spatial scales that can be precisely studied with a MHD description) was suggested by \citet{zhou90}. This model has been successively applied by \citet{li01} for addressing the transition between inertial and dissipation ranges. Their conclusion was that any kinetic linear damping mechanism would not be able to reproduce the observed power spectra of magnetic fluctuations, because it would produce a steep cut off instead of a power law, for high wavenumbers.\\
We will show that, by approaching the linear theory with a non-modal formalism, the diffusive model proposed by \citet{li01} does not necessarily rule out the possibility that a completely linear damping mechanism might be responsible for the ultimate dissipation of turbulent fluctuations. Moreover, we argue that the results based on a non-modal approach clarify the relationship between different eigenmodes, and clearly establish the fact that the problem of turbulent dissipation is very unlikely to be understood by studying the damping properties of one single normal mode, as previous works attempted to do. We will argue that, at kinetic scales, the evolution of a small perturbation depends very much on the linear interactions between many different modes, including heavily damped ones. Also, from a purely computational point of view, we will show that our results question the possibility of exciting one single eigenmode.\\

\subsection{Modal and non-modal approaches}
Plasma stability theory is traditionally treated as a normal mode (`modal') analysis. This means that the focus is on the eigenmodes of a slightly perturbed equilibrium, which are  assumed to grow or damp exponentially in time. The most rapidly growing (or the least damped) eigensolutions are the object of greatest interest in the normal mode analysis, which therefore investigates the stability problem purely in its time-asymptotic solutions, regardless of the details of the perturbation imposed on the initial equilibrium.\\
The non-modal approach differs from the normal mode analysis in treating the linear stability as an initial value problem. This allows the study of transient phenomena and to follow the evolution of a perturbed system in time. This evolution depends, of course, on how the initial equilibrium is perturbed.\\
The non-modal approach to the plasma stability problem can in some cases be crucial to the understanding of the evolution of a plasma in the linear regime, and it should therefore routinely be used. Especially when studying a system in stable conditions the normal-mode analysis misses the phenomenon of transient growth in time of some physical quantities. This is a well-known effect which has been long studied in hydrodynamic flows \citep{schmid07}. It is related to the spectral properties of the linear operator that describes the evolution of the system in time. In particular, when an operator is non-normal, i.e. it does not commute with its adjoint, the norm of an initial perturbation applied to the equilibrium can grow in time by large factors, before decaying, even if all the eigenmodes of the operator are damping (i.e. the equilibrium is stable).\\
Despite the fact that transient growth has been known of for a long time \citep{trefethen93}, it has not been emphasised in the theory of plasma stability. We have conducted the first study of transient growth for a stable kinetic plasma in a previous paper \citep{camporeale09}, showing that this phenomenon appears to be related to the kinetic regime of a plasma: it is indeed more accentuated for higher plasma beta and shorter spatial scales. In order to include kinetic effects we used a Landau fluid model, which incorporates linear Landau damping, and finite Larmor radius corrections. \\
\subsection{Aims of the paper}
The purpose of this paper is threefold. First, we extend the work of \citet{camporeale09} to an anisotropic plasma, for conditions typical of the solar wind. By doing so we will highlight the inadequacy of the normal mode analysis for a stable kinetic plasma.\\
Second, we will clarify the relation between transient growth and marginal stability, and we will propose a physical picture where the two observational contradictions mentioned above will be reconciled with the linear theory.\\
Third, we will discuss the importance and appropriateness of a non-modal linear theory for the understanding of the dissipation of turbulent fluctuations. It should be the goal of a future complete theory of solar wind turbulence to elaborate a unified theoretical framework that links  the physics of kinetic instabilities, and the damping of turbulent fluctuations in a consistent manner, and we believe that such a theory will benefit from embracing a non-modal approach to the stability problem, such as the one presented in this paper.\\
The paper is organised as follows.\\
In Section 2 we describe briefly the Landau fluid model, that has been used to obtain the results presented in the paper. The model has been described and commented in length elsewhere \citep{passot06,passot07,sulem08}, hence we will just briefly present the main features of the model, and we report the set of equations in the Appendix. In Section 3 we will introduce the methodology of the non-modal approach, and we will define some important quantities for our analysis. Section 4 describes the results of our study, with emphasis on the relationship between transient growth and marginal stability, and on the non-modal approach for the study of turbulent dissipative fluctuations. A final discussion, with suggestions for future work, is reported in Section 5.

\section{The FLR-Landau fluid model}\label{LF}

The idea of incorporating Landau damping in a set of fluid equations
was introduced by \citet{HP90} and later developed
in \citet{SHD97}, where the fluid hierarchy
obtained from the drift kinetic equation is closed at the level of the third
or fourth order moment. These models are limited to scales large
compared with the ion gyroradius. Other models, called gyrofluid models, consider the
fluid hierarchy obtained from the gyrokinetic equation, providing a
set of equations for fluid moments suitable for the description of sub-Larmor
radius scales \citep{Briz92,DH93,BH96}.
The equations of gyrofluid models, however, are not written in the physical coordinates 
but in the gyrocenter variables, making their interpretation more difficult.
A simpler formulation retaining hydrodynamic nonlinearities together with a linear
approximation of FLR contributions  was recently
developed  by  deriving equations for the hydrodynamic moments
directly from the Vlasov-Maxwell system \citep{GPS05,sulem08,passot07}. This is the model used in this paper. 
The hierarchy of fluid equations is closed at the level of the fourth order moment. In its linearized version, the plasma dispersion relation is approximated by a suitable Pad\'{e} approximant, and this allows to cast the linear set of equations as a standard eigenvalue problem. In addition to the 
hierarchy closure, this approach involves the modeling  of FLR
effects  by expressing  the non-gyrotropic 
part of tensors such as pressures, heat fluxes, or fourth order
moments in terms of lower-rank moments, in  a way consistent with the linear
kinetic theory in the low-frequency limit $\epsilon \sim \omega / \Omega_i\ll 1$, for
both   quasi-transverse fluctuations  ($k_\|/k_\perp \sim \epsilon$)  
with  no condition on $k_\perp r_L $ (as in gyrokinetic and gyrofluid
approaches),  but also for hydrodynamic  
scales with  $k_\| \sim k_\perp \ll 1/r_L$. Here $\Omega_i$ denotes
the ion cyclotron frequency and $r_L$ the ion Larmor radius.  
At large scales, the model,
which then reduces to usual anisotropic MHD, also captures the fast
waves, in contrast with gyrofluids. The frequency and damping rates of low-frequency waves
are accurately described in a range of scales that extends to small
(sub-Larmor radius) scales when the propagation direction is almost 
perpendicular to the ambient magnetic field (according to the
gyrokinetic scaling). The complete model is quite involved and is thoroughly described in
\citet{passot07}. The full set of equations is reported in Appendix, for completeness. The total number of variables is 16, hence the linear problem reduces to the formulation of a $16\times16$ complex matrix.

\section{Non-modal approach}\label{non-modal}
In this section we introduce the mathematical tools and the methodology that will be used in the rest of the paper. For a more complete description of the non-modal stability theory (in the context of hydrodynamics) we refer to the review by \citet{schmid07}; several issues related to the non-normality of a linear operator (that will be a central point of our discussion) have been described in great depth in the monograph by \citet{trefethen_book}.\\
The Landau fluid (LF) model can be linearized as usual, by writing each physical quantity as a sum of an equilibrium and a perturbed contribution (subscript 0 and 1 respectively): $\phi=\phi_0+\varepsilon\phi_1$, assuming that $\varepsilon\ll 1$, and neglecting terms of order higher than one in $\varepsilon$. Once the first-order variables are Fourier decomposed (dropping the subscript) $\phi(\mathbf{r},t)=\tilde{\phi}\exp[i(\mathbf{k\cdot r})]$, the linear LF model can be cast as a set of ordinary differential equations for the complex amplitudes $\tilde{\phi}$:
\begin{equation}\label{lin_set}
\frac{d\tilde{\phi}}{dt}=\mathbf{A}\tilde{\phi}(t).
\end{equation}
$\mathbf{A}$ is an autonomous (i.e. not a function of time) operator, that takes the form of a 16x16 sparse complex matrix, whose entries depend on the properties of the plasma (protons and electrons temperature anisotropy and plasma beta), and on the magnitude and angle of propagation of the wavevector $\mathbf{k}$.\\
The fact that the LF model describes the plasma through a set of only 16 variables makes the analysis of the matrix $\mathbf{A}$ computationally affordable without any particular method used for large matrix manipulations. Accordingly, all the results presented in this paper have been produced using in-built routines of MATLAB.\\
The solution of Eq.(\ref{lin_set}) is given by
\begin{equation}
\tilde{\phi}(t)=e^{\mathbf{A}t}\tilde{\phi}(0),
\end{equation}
where $\tilde{\phi}(0)$ is the state vector of the initial perturbation and the exponential of the matrix (which is defined as $e^{\mathbf{A}t}=\mathbf{I}+\mathbf{A}t+\frac{1}{2}(\mathbf{A}t)^2+\ldots$) completely determines the evolution in time of the initial state (it is sometime called the propagator).\\
A distinctive feature of the matrix $\mathbf{A}$ that is crucial to our argument is its normality. If $\mathbf{A}$ commutes with its adjoint $\mathbf{AA^*}=\mathbf{A^*A}$, than $\mathbf{A}$ is said to be normal, and its eigenvectors form a complete orthogonal set. We note that the definition of normality depends on which norm one refers to. Any non-normal linear operator can be made normal by choosing a different definition of the norm, from which the definition of adjoint follows \citep{farrell96}. We use here the 2-norm $\|u\|=\sqrt{\sum_{i=1}^{16}|u_i|^2}$. Altough this norm does not provide an information on single variables, it indicates whether the perturbation complies or not to the requirement of being much smaller than the equilibrium quantities. What is important is that a large norm of a perturbation implies a deviation from the assumption of linearity, and therefore might result in the triggering of non-linear effects.\\
\subsection{Spectral abscissa and numerical abscissa}
In order to study how the plasma responds to a small perturbation we introduce the quantity $G(t)$, which measures the amplification (or reduction) in time of an initial perturbation:
\begin{equation}
G(t)=\frac{\|\tilde{\phi}(t)\|}{\|\tilde{\phi}(0)\|}=\frac{\|e^{\mathbf{A}t}\tilde{\phi}(0)\|}{\|\tilde{\phi}(0)\|}.
\end{equation}
This quantity clearly depends on the details of the initial perturbation $\tilde{\phi}(0)$, but it is always bounded by above from the quantity $\|e^{\mathbf{A}t}\|$, which defines the supremum of $G(t)$ for any possible $\tilde{\phi}(0)$. The behaviour of $\|e^{\mathbf{A}t}\|$ depends on the spectral properties of $\mathbf{A}$. If one indicates with $\sigma(\mathbf{A})$ the spectrum of $\mathbf{A}$, i.e. the set of $z\in\mathbb{C}$ such that $(z\mathbf{I}-\mathbf{A})$ is singular, where $\mathbf{I}$ is the identity operator, then the quantity 
\begin{equation}
\alpha(\mathbf{A})=\max{\Re[\sigma(\mathbf{A})]}
\end{equation}
is referred to as the spectral abscissa. To determine this quantity is the main objective of the normal-mode analysis, because it gives the growth rate of the most unstable mode (for an instability), or the damping rate of the least damped mode (for a stable plasma). Hence, the time-asymptotic evolution of any initial perturbation is given by $e^{\alpha(\mathbf{A})t}$. Moreover, for a normal operator, the spectral abscissa bounds the quantity $G(t)$, for any time $t\geq 0$:
\begin{equation}\label{spectral_abs}
\sup\; G(t)=\|e^{\mathbf{A}t}\|=e^{\alpha(\mathbf{A})t}\;\;\mbox{(for $t\geq0$   iff $\mathbf{A}$ is normal)}.
\end{equation}
In general however, for a non-normal operator, Eq.(\ref{spectral_abs}) holds only in the limit $t\rightarrow\infty$, and there is no exact formula for the quantity $\|e^{\mathbf{A}t}\|$, that can only be approximately estimated for $t>0$ (different approximations can be found in \citet{trefethen_book}).\\
But a rigorous formula exists for the growth of $\|e^{\mathbf{A}t}\|$ at $t=0$. This is referred to as the numerical abscissa:
\begin{equation}\label{numerical_abs}
\eta(\mathbf{A})=\left\vert\frac{d}{dt}\|e^{\mathbf{A}t}\|\right\vert_{t=0}=\sup\sigma\left(\frac{1}{2}(\mathbf{A}+\mathbf{A^*})\right).
\end{equation}
Note that since $(\mathbf{A}+\mathbf{A^*})$ is Hermitian, $\eta(\mathbf{A})$ is real. The numerical abscissa provides the information about the highest possible growth rate of any initial perturbation, at time $t=0$. For a normal operator it follows from Eq.(\ref{numerical_abs}) that $\eta(\mathbf{A})=\alpha(\mathbf{A})$, consistently with the fact that, if $\mathbf{A}$ is normal, the quantity $G(t)$ is bounded by $e^{\alpha(\mathbf{A})t}$ for any time.\\
What is surprising is that, for a non-normal operator, even if all the normal modes are damping (i.e. $\alpha(\mathbf{A})<0$) the numerical abscissa can be positive, hence allowing the quantity $G(t)$ to grow for some particular initial perturbations. It has been shown in \citet{camporeale09} that $G(t)$ can indeed reach values of about $10^3-10^4$ over short periods of time, for a Maxwellian plasma described by the Landau fluid model.\\
The fact that a perturbation could grow while the eigenmodes of the operator decay in time appears counter-intuitive. It is purely due to the superposition of non-orthogonal eigenvectors, and we refer to Figure 2 in \citet{schmid07} for a graphical paradigmatic explanation of this effect.
\subsection{Pseudospectra}
\begin{figure}
 \center\includegraphics[width=25pc]{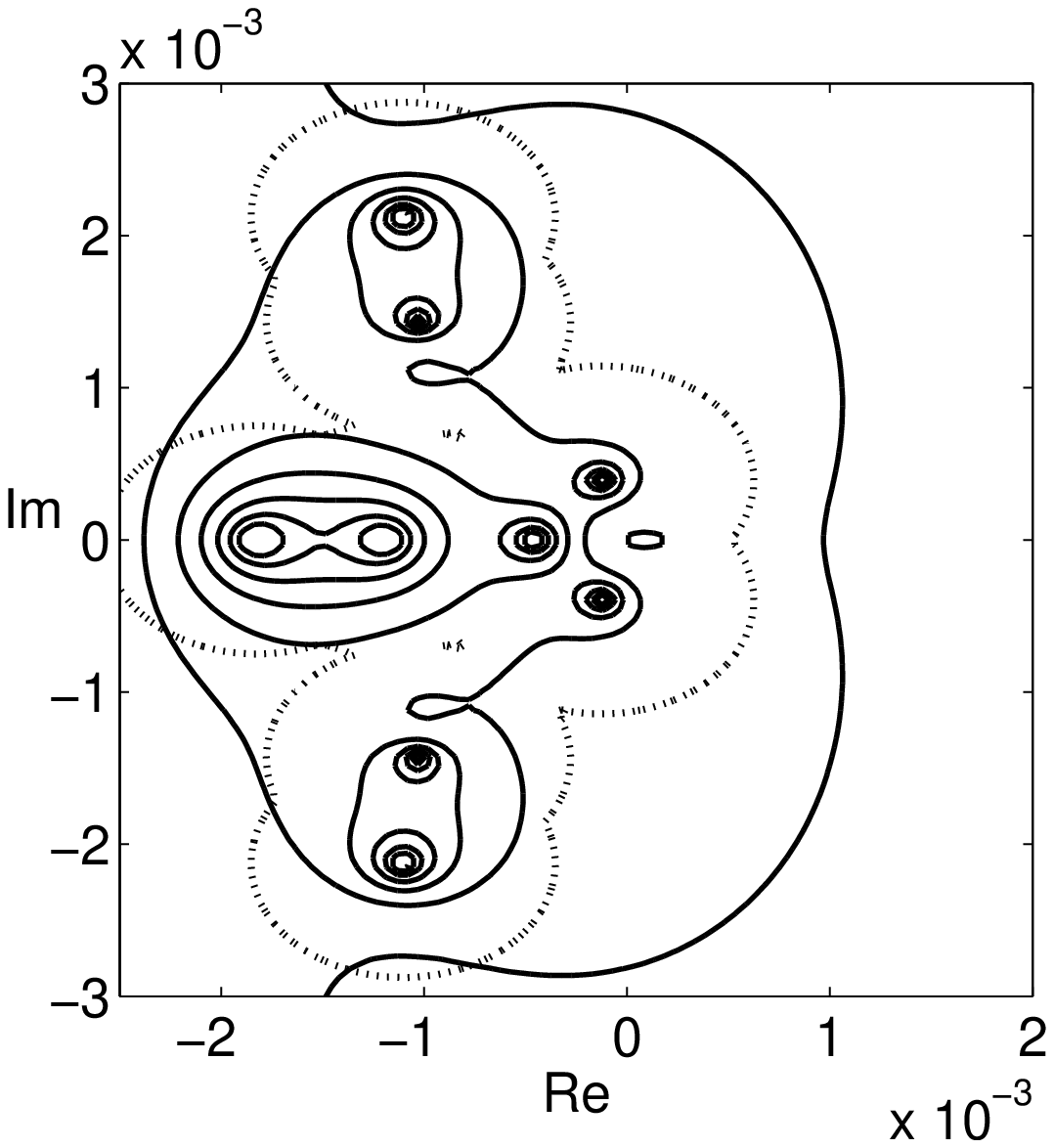}
 \caption{Contours of the $\varepsilon$-pseudospectrum for a plasma with $\beta=10$, $k=0.5$, $T_\perp/T_\parallel=1$. Contours are plotted for $log_{10} \varepsilon = -6.8, -6.6, \ldots, -5.6$. The dotted line is how the $\varepsilon$-contour would appear if the operator were normal, for $\varepsilon=0.00075$.}\label{fig1}
 \end{figure}

A key  aspect of non-normal operators is that the spectrum may be highly sensitive to small perturbations. In order to quantify the effect of small perturbations on a linear operator, the concept of pseudospectra has been introduced. There are at least four definitions of pseudospectra, that have been shown to be mathematically equivalent \citep{trefethen_book}. We report here the two that have a most immediate link with a physical interpretation. If one employs the convention that the spectrum $\sigma(\mathbf{A})$ (the set of eigenvalues) is formed by complex values $z$ for which $\|(z\mathbf{I}-\mathbf{A})^{-1}\|=\infty$, the $\varepsilon$-pseudospectrum $\sigma_\varepsilon(\mathbf{A})$ is defined as the set of $z\in\mathbb{C}$ such that
\begin{equation*}
\|(z\mathbf{I}-\mathbf{A})^{-1}\|>\frac{1}{\varepsilon}\;\;\;\;\mbox{(Definition 1)}
\end{equation*}
for any $\varepsilon>0$.\\
This is equivalent to saying that $z$ is an eigenvalue of the perturbed operator $(\mathbf{A}+\mathbf{E})$ for some operator $\mathbf{E}$ with $\|\mathbf{E}\|<\varepsilon$, i.e.:
\begin{equation*}
z\in\sigma_\varepsilon(\mathbf{A}) \Longleftrightarrow z\in\sigma(\mathbf{A}+\mathbf{E})\;\;\;\;\mbox{with $\|\mathbf{E}\|<\varepsilon$    (Definition 2)}
\end{equation*}
Let us now comment on these definitions and their meaning in a more physical sense. Our linear operator studies the evolution of a small perturbation applied to an equilibrium, namely a bi-Maxwellian plasma in an uniform magnetic field. However, one could argue that small discontinuities or inhomogeneities of the equilibrium quantities might be modelled as small disturbances to the linear operator. The question that pseudospectra quantitatively answer is: how does the set of eigenvalues change under the effect of small perturbations of the operator ? From definitions 1 and 2, one see that the $\varepsilon$-pseudospectrum $\sigma_\varepsilon$ tends precisely to the standard spectrum $\sigma$, when $\varepsilon\rightarrow 0$. In the complex plane the $\varepsilon$-pseudospectra are the open subset that contains all the eigenvalues of the perturbed operator $(\mathbf{A}+\mathbf{E})$, for any possible perturbation $\mathbf{E}$, such that $\|\mathbf{E}\|<\varepsilon$. Therefore they give a measure of the distortion of the spectrum due to the perturbation applied to the operator. It is straightforward to prove that the $\varepsilon$-pseudospectra are a nested set. That is, $\sigma_{\varepsilon_1}(\mathbf{A})\subseteq\sigma_{\varepsilon_2}(\mathbf{A})$ for $\varepsilon_1\leq\varepsilon_2$\\
An example of pseudospectra of our LF operator is given in Figure \ref{fig1}, for an isotropic plasma with $\beta=10$ and $k=0.5$. Each contour corresponds to values of $\varepsilon=10^{-6.8},10^{-6.6},\cdots,10^{-5.6}$. Only a part of the complex plane is shown, with 9 of the 16 eigenvalues visible. The important point here is to understand that perturbations of a certain size make the $\varepsilon$-pseudospectrum so large, that it would contain many eigenvalues of the original unperturbed operator. In a sense, this means that such perturbations distort the problem in such a way that the information about the modes of the unperturbed operator becomes useless. In this respect the different behaviour between normal and non-normal operator is most evident. The $\varepsilon$-pseudospectrum of a normal operator is defined as the union of open balls about the points of the spectrum:
\begin{equation*}
\|(z\mathbf{I}-\mathbf{A})^{-1}\|=\frac{1}{\mathrm{dist}[z,\sigma(\mathbf{A})]}
\end{equation*}
where dist indicates the distance of a point to a set in the complex plane. For a normal operator the curves of the pseudospectra can be computed straightforwardly once the eigenvalues are known. We show in Figure \ref{fig1}, with dotted lines, how the contours of $\sigma_\varepsilon$ would be for the same LF operator, if it were normal, for $\varepsilon=0.00075$. This contour is qualitatively similar to the (true) contour for $\varepsilon=10^{-5.8}$, embracing all the 9 eigenvalues. This is the crucial point about the different behaviour between normal and non-normal operators. A perturbation of the order $~10^{-5.8}$ is sufficient to achieve a distortion of the spectrum that, if the operator would have been normal, would have required a perturbation of the order $~7.5\cdot 10^{-4}$, i.e. about 470 times higher.\\
The displacement of the spectrum of  an operator subject to small perturbations implies two facts. On one hand some, if not all, of the modes become somehow coupled, i.e, their relative distance in the complex plane can be modified, and their properties (like phase speed and damping rate) changed. On the other hand, from a computational point of view, the non-normality of the operator almost completely rules out the possibility of exactly exciting one single eigenvector. Small errors, due for instance to digits truncation or approximation, can result in the excitation of a `pseudomode', that could lie in the complex space, very far from the mode that was intended to be excited, and that is in reality a superposition of different non-orthogonal modes. This might lead to an initial transient behaviour, which is not unusual in numerical simulations, even though it is seldom commented.

\section{Results}\label{results}
In this section we apply the linear LF model to a stable proton-electron ($p,e$) bi-Maxwellian plasma in an uniform magnetic field. We are interested in how the distance from marginal stability affects the evolution of linear perturbations in the plasma, and wish to address transient behavior that is missed by the standard modal analysis. For all the results shown in this section the protons are considered isotropic with  $T_{\perp p}=T_{\parallel p}=T_{\parallel e}$. All quantities are therefore referred to electrons, and the subscript $e$ is dropped.\\
We focus on quasi-perpendicular waves, since the LF model has a domain of validity that extends to small scales only for oblique wavevectors. The angle of propagation for the wavevector $k$ with respect to the background magnetic field is $\theta=\tan^{-1}(1000)$. \\
It is known that the marginal stability thresholds (i.e. the curves for which the growth rate is exactly null) obey a law of the form: $\frac{T_\perp}{T_\parallel}=1+\frac{S}{\beta^\alpha}$, where $S$ and $\alpha$ are constants. We have derived the stability thresholds for the LF model, and we have computed the fitting parameters using a Levenberg-Marquardat method \citep{press}. The result is:
\begin{equation}\label{marg1}
\frac{T_\perp}{T_\parallel}=1-\frac{1.9884}{\beta}\;\;\;\mbox{(for $T_\parallel>T_\perp$)}
\end{equation}
and
\begin{equation}\label{marg2}
\frac{T_\perp}{T_\parallel}=1+\frac{0.88}{\beta}\;\;\;\mbox{(for $T_\parallel<T_\perp$)}
\end{equation}
The values for $T_\parallel>T_\perp$ are not dissimilar from those obtained from the Vlasov-Maxwell equations for the electron firehose instability ($S=-1.29, \alpha=0.98$), which is known to yield highest growth rate for quasi-perpendicular propagation \citep{camporeale08}.  The parameters for $T_\parallel<T_\perp$ are instead not comparable with those obtained for the mirror and whistler instabilities \citep{gary96}, because in this case the whistler instability, which is parallel propagating, sets the instability threshold \citep{gary06}, and the Landau fluid model becomes invalid for strictly parallel wavevectors. We therefore will focus mainly on the case $T_\parallel>T_\perp$.\\
\subsection{Transient growth and marginal stability}
\begin{figure}
 \center\includegraphics[width=20pc]{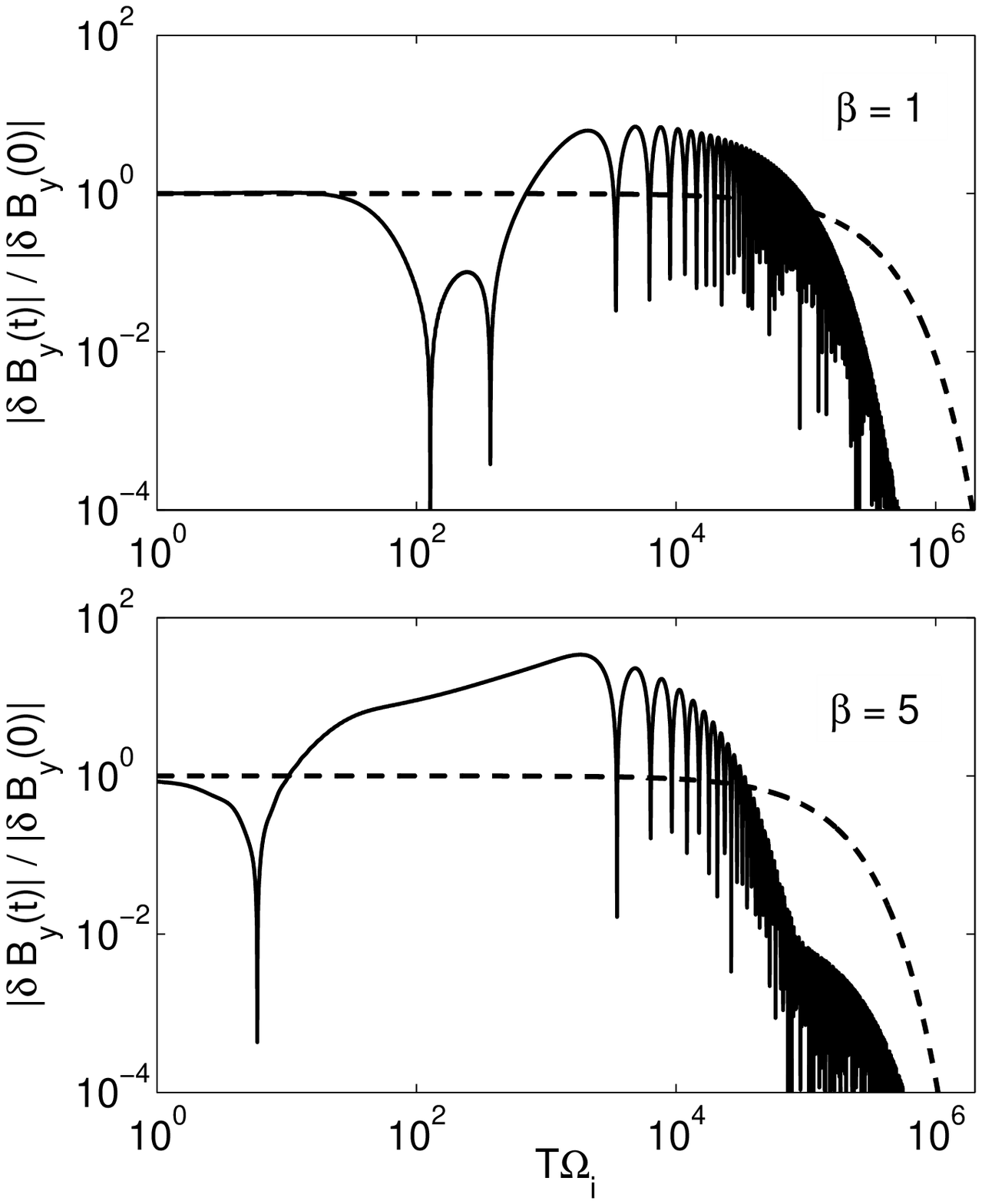}
 \caption{Evolution in time of the absolute value of the amplitude of the $y$ component of the magnetic field, normalized to its initial value, for one particular choice of initial condition. The parameters used are: $k=1$, $T_\perp/T_\parallel=1$, $\beta=1$  (top panel) and $\beta=5$ (bottom panel). The curve in dashed line is the evolution $e^{\alpha t}$, predicted by modal theory.}\label{fig3}
 \end{figure}

\begin{figure}
 \center\includegraphics[width=20pc]{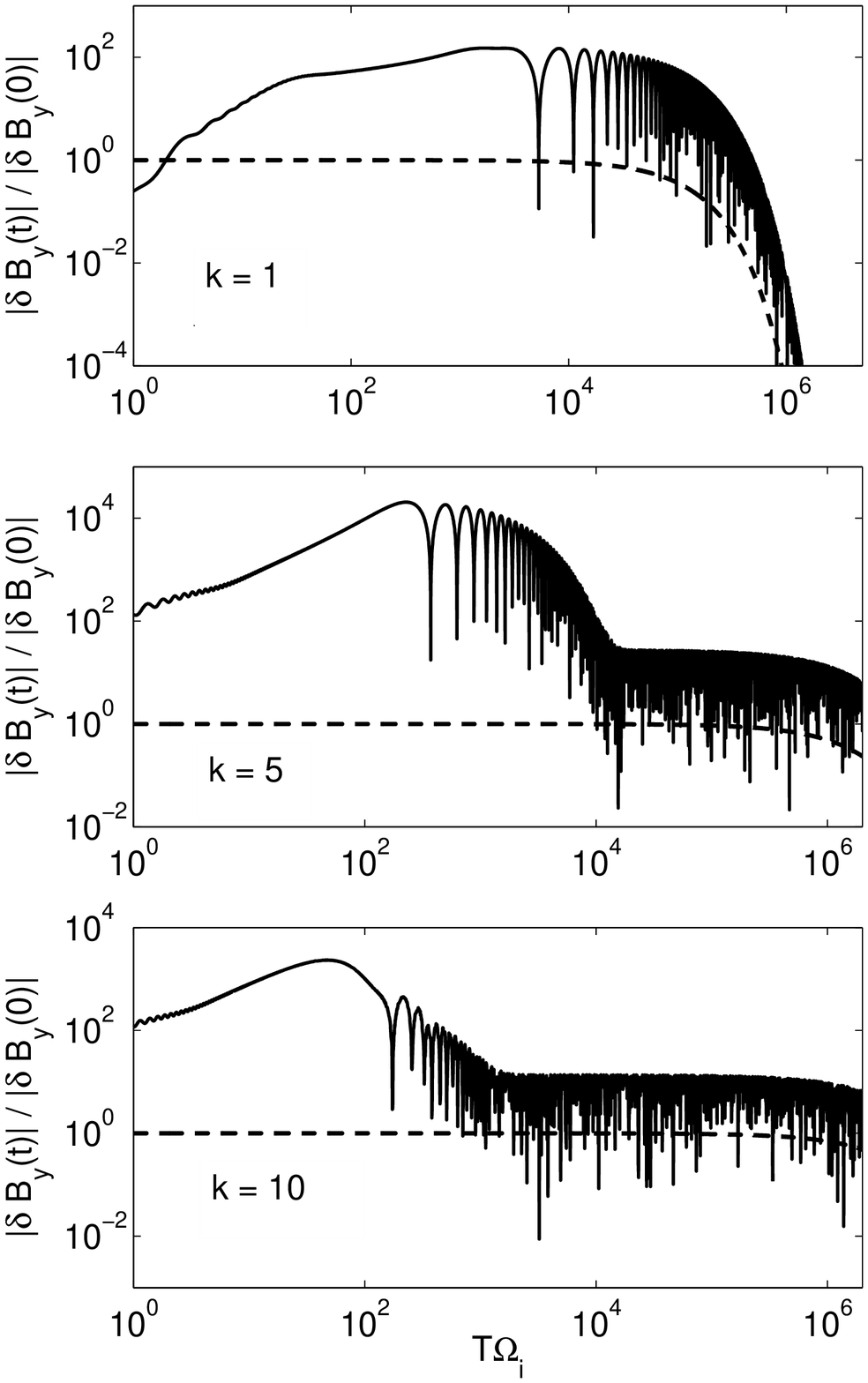}
 \caption{Evolution in time of the absolute value of the amplitude of the $y$ component of the magnetic field, normalized to its initial value, for one particular choice of initial condition. The parameters used are: $T_\perp/T_\parallel=0.65$, $\beta=5$, $k=1$  (top panel), $k=5$ (central panel), and $k=10$ (bottom panel). The curve in dashed line is the evolution $e^{\alpha t}$, predicted by modal theory.}\label{fig3b}
 \end{figure}

\begin{figure}
 \center\includegraphics[width=40pc]{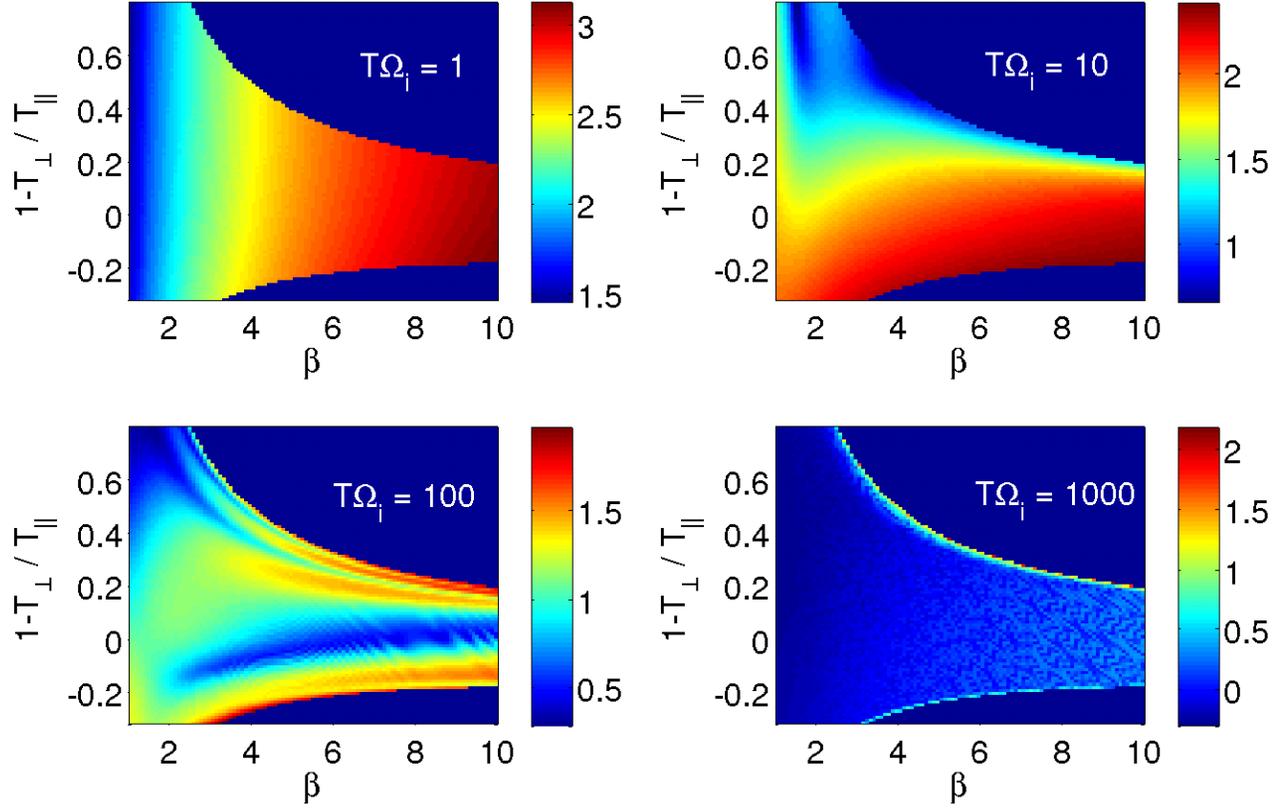}
 \caption{(Color online) Average value of G(t) computed over 10000 randomly generated initial perturbations, shown at four different times, in the $(\beta,1-T_\perp/T_\parallel)$ space. Logarithmic scale.}\label{mean_norm}
 \end{figure}

\begin{figure}
 \noindent\includegraphics[width=40pc]{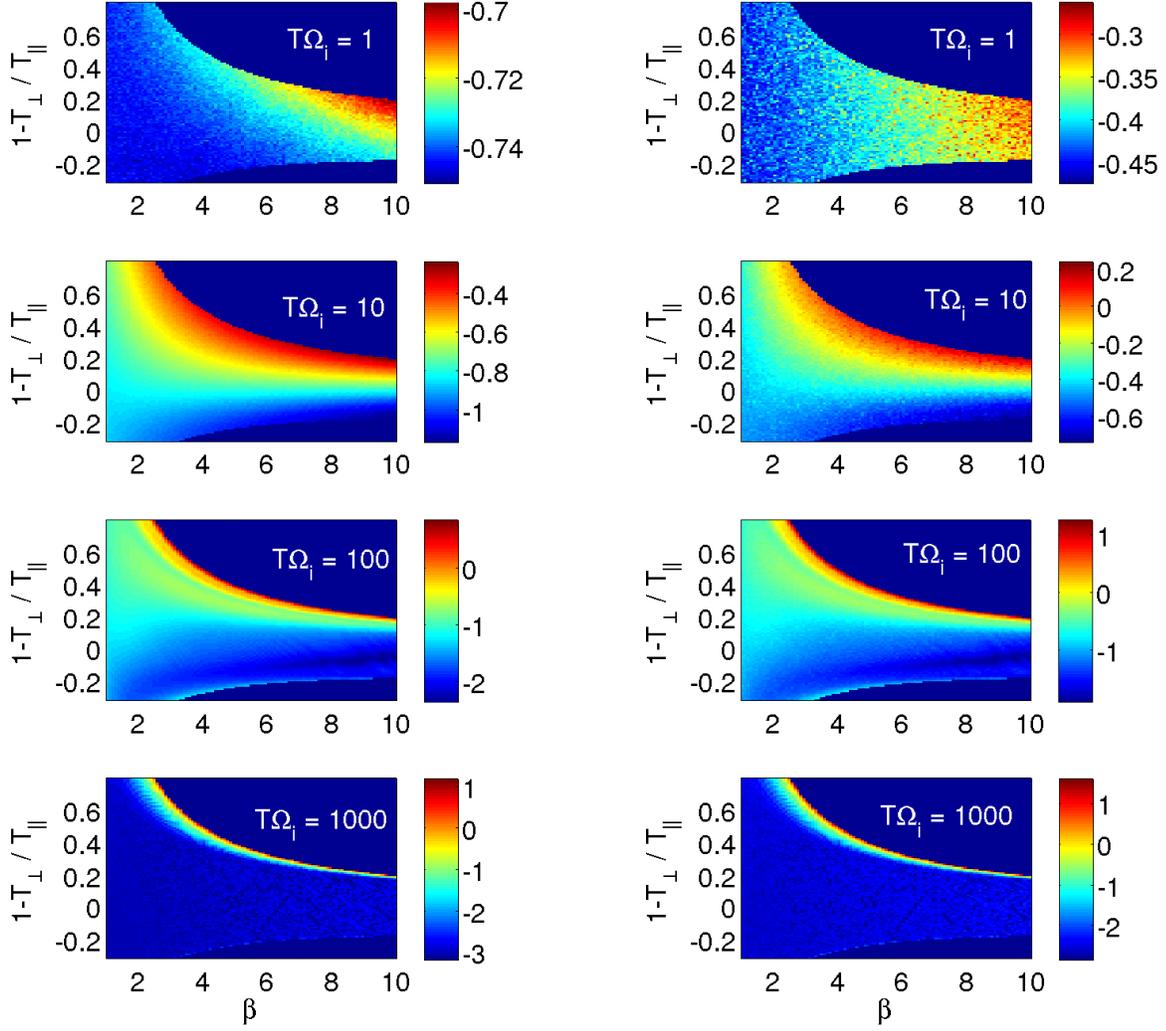}
 \caption{(Color online) Average (left panel) and maximum (right panel) values of $\delta B/B_0$ computed over 10000 randomly generated initial perturbations, shown at four different times, in the $(\beta,1-T_\perp/T_\parallel)$ space. Logarithmic scale}\label{delta_B}
 \end{figure}

It has been shown in \citet{camporeale09} that an isotropic Maxwellian plasma is able to sustain large transient growth of an initial perturbation, that will eventually decay in later times.\\
As we have seen in Section 3, the spectral abscissa $\alpha$ (which is negative, for a stable plasma) provides the information about the late-time damping rate of the perturbation. That is, any initial fluctuation will decay as $e^{\alpha t}$, for large times. An interesting point is to actually analyse the time at which the system starts to behave as predicted by the normal mode-analysis.\\
We show in Figure \ref{fig3} the amplitude of the $y$ component of the magnetic field, normalized to its initial value, for two particular initial conditions. The wavevector $k$ is chosen equal to 1, and $T_\perp/T_\parallel=1$. The parameter $\beta$ is equal to 1 in the top panel, and $\beta=5$ in the bottom panel. What emerges is that, for both cases, the behavior is highly oscillatory, and no damping is evident before $T\Omega_i=10^5$. This implies that, for the initial perturbation to decay, the plasma, once perturbed, should evolve without encountering any further perturbation for a very large time, which is quite unrealistic. The damping predicted by modal analysis (i.e. $e^{\alpha t}$) is shown in dashed line.\\
Before proceeding, let us clarify how the initial perturbations of Figure \ref{fig3} were chosen. Let us recall that the singular value decomposition (SVD) of an operator $\mathbf{A}$ is given by
\begin{equation}
\mathbf{A}=\mathbf{U\Sigma V^*},
\end{equation}
where $\mathbf{U}$ and $\mathbf{V}$ are unitary matrices, and $\mathbf{\Sigma}$ is a diagonal matrix, that contains the singular values of $\mathbf{A}$.\\
If one wants to find the initial perturbation $\tilde{\phi}(0)$, that at a specific time $\Theta$ undergoes an amplification $G(\Theta)$ equal to $\|e^{\mathbf{A}\Theta}\|$ (i.e. the highest possible amplification at time $\Theta$), it is sufficient to calculate the SVD of $e^{\mathbf{A}\Theta}$, and to identify the column vector of $\mathbf{V}$ associated with the highest singular value. The initial perturbations chosen to produce Figures \ref{fig3} and \ref{fig3b} are the ones that approximately attain the highest amplification at a certain time. One could argue that these initial conditions are very special, and cannot represent the totality of all the possible initial states. This is certainly true. However, it is not feasible to apply the non-modal approach to a very large set of different conditions (that will never be large enough to represent `all' cases). Beside, the methodology that we use, i.e., to focus on only certain particular cases that represent the `worst case scenario', is common also to the modal approach, whose results are supposed to be valid for any initial conditions (albeit restricted to asymptotic times), but still represent only the scenario in which the fastest growing (or the least damped) mode is actually excited. Later we describe an attempt to approach the problem in a more statistical manner.\\
Figure \ref{fig3} shows the results of two cases for an isotropic plasma, with different $\beta$, but what is perhaps more interesting is to look at what happens when the plasma is closer to marginal stability, given by Eqs. (\ref{marg1}) and (\ref{marg2}). For $\beta=5$, Eq. (\ref{marg1}) predicts that the plasma is marginally stable when $T_\perp/T_\parallel\sim 0.6$. We show in Figure \ref{fig3b} three cases of transient growth with $\beta=5$ and $T_\perp/T_\parallel=0.65$, for $k=1,5,10$. The spectral abscissa (damping rate) is respectively $\alpha=-1.02\cdot 10^{-5},-7.2\cdot 10^{-7},-3.9\cdot 10^{-7}$.\\
Since the damping rate is so low, what is expected is that any perturbation applied to the plasma would remain unchanged in amplitude for long times. This is indeed what is observed. What is rather interesting, however, is that the initial perturbations shown in Figure \ref{fig3b} undergo a transient growth at early times, and therefore remain amplified for long times. In other words, a transient growth effect is able to amplify a perturbation even at marginal stability condition, and the fact that the damping rate is very small allows to the amplified perturbation to survive for long times. In the examples of Figure \ref{fig3b} the initial value of $B_y$ gets amplified by a factor between 10 and 100 and remains so at least for $10^5$ ion gyroperiods.\\
We now try to quantify the importance of transient growth effects in a statistical sense for the following range of parameters: $\beta\in[1.5,10],T_\perp/T_\parallel\in[0.2,1.2]$.\\
We have divided the $(\beta,T_\perp/T_\parallel)$ space in a $100\times 100$ grid. For each point in the stable region we have generated 10000 random initial perturbations. In Figure \ref{mean_norm} we show the mean value of $G(t)$ at times $t=1,10,100,1000$, in $\log_{10}$ scale. We recall that $G(t)$ measures the amplifications of the 2-norm on the vector, hence considering all the 16 variables equally. One can notice that at time $t=1$, $G(t)$ increases r for higher $\beta$, almost independently of the value of the anisotropy. Transient growth are therefore not a sporadic event, even for an isotropic plasma. For later times $G(t)$ has still large values, and the highest peaks tend to be localized at the edge of marginal stability, for increasing time. In figure \ref{delta_B} we show the average (left panels) and the maximum (right panels) value of $\delta B/B_0$ calculated across the 10000 different initial perturbations. The highest magnetic fluctuations are evident in proximity of the $T_\parallel>T_\perp$ instability threshold, for any time. Once again the transient growth of magnetic fluctuations appears not to be a sporadic and unusual event that happens only for carefully chosen initial perturbations, but a feature that is persistent in time.\\
In the light of these results we suggest the following scenario that might explain both the presence of fluctuations at high wavenumbers in stable conditions, and the fact that most of the solar wind is observed close to temperature isotropy, despite the expansion.\\
The increase of anisotropy is expected, in fluid theories, due to the conservation of one or more adiabatic invariants. For instance, the CGL approximation \citep{CGL} predicts that the ratio $T_\parallel/T_\perp$ increases as $r^2$ (where $r$ is the distance from the Sun), if the plasma is expanding radially in a magnetic field that varies as $r^{-2}$. Given that a temperature anisotropy is a source of free energy for instabilities, it has been argued that the rise of anisotropy should be possible only until the free energy becomes so large that an instability might be triggered, with the consequence of not allowing a further enhancement of anisotropy, and therefore to bound also the increase of free energy.\\
However, we have shown that, in stable conditions, small perturbations can give rise to transient growth of magnetic fluctuations and of high order moments of the particle distribution function. We argue that the presence of these fluctuations, even far from the stability threshold, can influence the ability of the plasma to further increase the free energy via the anisotropization of the particle distribution function. In fact, the thresholds for marginal stability are derived from the linear theory as the conditions for which the time asymptotic growth rate $\alpha$ becomes null, assuming for the plasma to be in equilibrium in an uniform magnetic field. No consideration is made for the fact that the growth rate is effectively a function of time, and that the magnetic field might not be uniform, due to transient fluctuations. What might happen is therefore that the increase of anisotropy might be bounded even before reaching what is considered the threshold for anisotropy instabilities. In this scenario, a parcel of plasma could experience a `local' marginal stability condition due to a temporary enhanced magnetic fluctuation.\\
This scenario reconciles the observational contradictions mentioned above with a consistent interpretation of the linear theory. Of course, trying to interpret space observations purely on the base of linear theory might appear a naive approach and a complete and coherent understanding of the process of the solar wind expansion is feasible only via a non-linear treatment, and through computer simulations. Howewer, our results show that a non-modal linear theory could already constitute a strong theoretical ground for data interpretation.

\subsection{The dissipation of turbulent fluctuation}
\begin{figure}
 \center\includegraphics[width=20pc]{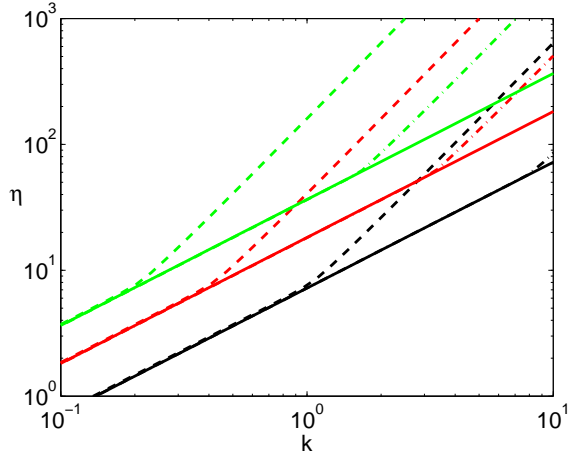}
 \noindent\caption{Numerical abscissa $\eta$ as a function of $k$ for the following parameters: $\beta=2$ (black), 5 (red), 10 (green), and for $T_\perp/T_\parallel=0.5$ (dot-dashed), $1$ (solid line), $2$ (dashed). Note that the isotropic case is marked with solid line.}\label{fig4}
 \end{figure}

In this section we highlight the particular behavior of the numerical abscissa as a function of wavenumber, and point out certain unique features which indicate that nonmodal effects may play an important part in the cascade of turbulent fluctuations at short wavelengths.\\
An approach that is very often used in modelling the dissipation at short wavelength of turbulent fluctuations is to assume that such small fluctuations can be regarded as an ensemble of linear waves, each of them damping according to their linear damping rate. \citet{howes08a} have devised a gyrokinetic formalism to study turbulence in a magnetized plasma, in the framework of the Goldreich and Sridhar critical balance assumption \citep{sridhar94}. They have shown, by performing nonlinear simulations, that, in the case of small damping rates, the linear damping does not underestimate the rate at which electromagnetic energy is dissipated \citep{howes08b}. However, the gyrokinetic cascade model produces an exponential roll-off of the power spectrum for high wavenumbers, instead of the observed power law \citep{leamon98a}. The fact that a linear damping mechanism cannot reproduce a power-law in the spectrum, for high wavenumbers, was previously pointed out by \citet{li01}, using the full Vlasov-Maxwell linear theory. Our contribution here is to revisit the conclusions reached by \citet{li01}, by showing that a non-modal approach might be more appropriate to address this problem, which is still largely unsolved.\\
The equation used by \cite{li01} to model the spectral energy $W(k)$ in wavenumber space is of the kind:
\begin{equation}
\frac{\partial W(k)}{\partial t}=D(k)W(k) + \gamma(k)W(k)+S(k),
\end{equation}
where $D(k)$ is a diffusion operator (that can be modeled in such a way to recover a Kolmogorov cascade for large wavelengths), $S(k)=S(k_0)\delta(k-k_0)$ is a source term that operates at the (small) injection wavenumber $k_0$, and $\gamma(k)$ is the damping rate given by the Vlasov-Maxwell linear theory. This model is clearly oversimplified in at least three ways. First, it assumes that the nonlinear cascade is a diffusive process in wavenumber space. Second, it assumes that the cascade is isotropic in $k$ (while it is known that anisotropy plays a very important role). Third, it describes the damping process entirely via an ensemble of independent linear waves, all propagating in the same direction, and each damping according to its respective linear damping rate.\\
Despite the weaknesses of the model, it still represents a good starting point for our purpose of highlighting the importance of transient, non-asymptotic dynamics. The main conclusion of \citet{li01} was that `a power-law spectrum in $W(k)$, in the presence of damping, requires $\gamma(k)$ to be a specific power-law'. However, at the light of the considerations hitherto drawn, we point out that treating the cascade process as a time-asymptotic problem rather than an initial value problem, hugely distorts the physics. In fact, each wavenumber is continuously subject to injection of energy via the non-linear cascade, and to treat the evolution of a fluctuation, as if it was injected in a remote past, and let free to damp undisturbed is, in our view, an oversimplification of the whole process.\\
Although it has been shown that some fluid models are able to recover the steepening of the power spectrum purely by a nonlinear mechanism (see e.g. \citet{galtier07}), we note that if one assumes for the non-linear terms in the cascade to act smoothly as a function of $k$, then it is very likely that at least one of the parameters that control the damping at high frequencies must present  some sort of abrupt change in their dependence on $k$.\\
This issue necessitates a deeper and separate study, and we do not intend to suggest here a definitive answer. However, it is interesting to show the behaviour of the numerical abscissa, as a function of $k$, for different values of $\beta$, and $T_\perp/T_\parallel$ (Figure \ref{fig4}). We recall that the numerical abscissa measures the derivative of the highest amplification that an initial perturbation can undergo, at time $t=0$.\\
Two features are extremely surprising, in Figure \ref{fig4}. First, the numerical abscissa $\eta$ follows a power-law in wavenumbers (i.e. a straight line in the log-log plot). Second, it presents an abrupt steepening of its slope, when $T_\perp/T_\parallel\neq 1$. As far as we know, this is the only quantity derived from the linear theory that presents this two features. Therefore, it makes sense to suggest that the numerical abscissa should be included in the set of parameters that control the cascade of turbulent fluctuations at short wavelengths. In some sense, this strengthen our argument that a non-modal approach to the problem might be decisive, and we leave further considerations for the future.

\section{Discussion}\label{discussion}
Plasma linear theory is traditionally carried out as a normal mode analysis. The modal approach, although useful to define the conditions for which an equilibrium might be unstable, is, in general, unable to describe the time evolution of some initial perturbations. In particular, it is known that some linear operators can present transient growth of physical quantities, even in stable conditions. We have addressed the stability of a kinetic plasma modelled through a Landau fluid model, via a non-modal approach. The following results have been established.
\begin{itemize}
\item  Some small initial perturbations can undergo large amplifications, even if applied to a stable equilibrium (Figure \ref{fig3}). If the plasma is close to marginal stability, this might result in a persistent presence of such fluctuations (Figure \ref{fig3b}). This is because the (time-asymptotic) damping rate is so small, that the decay of an initial perturbation is appreciable only after a time of the order of $\sim10^6$ ion gyroperiods. On the other hand, it is quite unrealistic for a plasma to be undisturbed for such large times.
\item  Transient growth of magnetic fluctuations, or of high order moments of the distribution function are quite possible also for nearly isotropic plasma, i.e far from the thresholds of linear anisotropy instabilities (Figures \ref{mean_norm} and \ref{delta_B}).
\item  The numerical abscissa, which is the highest growth rate, at time $t=0$, for any possible perturbation turns out to follow a power-law in $k$, and to present an abrupt change in slope for high wavenumbers.
\end{itemize}
We believe that these results could help to interpret some space observations in a more consistent way, than the traditional modal theory is able to do. We suggest that marginal fluctuations observed in stable conditions might be the result of transient growth. We also suggest that the marginal stability concept should be revisited within a non-modal context, and that the prediction of the evolution of small fluctuations in a turbulent plasma purely by their time-asymptotic damping rate, is an oversimplification of the physics.\\
Most of the solar wind protons and electrons are observed to be nearly isotropic. This is in contradiction with fluid theories that predict a continuous anisotropization of the distribution function. We suggest that the increase in anisotropy might be inhibited by the presence of ongoing transient effects, that could limit the capability of the plasma to increase the free energy.\\
As for the dissipation of turbulent fluctuations, it is now acknowledged that the understanding of this complex phenomenon, lays in the realm of kinetic plasma physics. However, once again the modal approach does not seem to be very useful here, and might be misleading. On the other hand, the power-law dependence of the numerical abscissa, and more importantly the presence of a steepening for high wavenumbers, is very interesting. We have pointed out that, since this is the only quantity (derived within linear theory) that presents such features, it should be taken into account.\\
\subsection{Future directions}
The primary intent of this paper has been to show that a non-modal approach could be helpful in reconciling some observational evidence with plasma linear theory, and to suggest a more frequent use of such an approach. Historically, the issues that can arise for a non-normal linear operator have been studied in depth in hydrodynamics, but have not captured the same attention in the space plasma community.\\
This of course does not exclude the use of non-linear theory and computer simulations, that will ultimately be the only way to fully understand the processes in action. However, we have shown that the interpretation of simulations should be made with the awareness of the effects produced by non-normal linear operators.\\
Altough the Landau fluid model used in this paper is superior to MHD, being able to capture some kinetic effects, it will be desirable in the future to develop a non-modal treatment of the full Vlasov-Maxwell set of equations.\\
Also, as we have shown, the non-modal approach should be framed in a more statistical theoretical framework.\\
In conclusion, we believe that the understanding and the ability to address transient behaviour will be a crucial ingredient in future modelling of space plasma.

%
%

\begin{acknowledgments}
This work was partially supported by STFC grant PP/E001424/1 and by ``Programme National Soleil Terre'' of CNRS. 
\end{acknowledgments}

\appendix

\section{FLR-Landau fluid set of equations}

In the context of this paper, it is useful to specify the form of 
the model in one space dimension, assuming that all the fields 
only depend on a coordinate $\xi$, along a direction of the $(x,z)$-plane making an angle
$\alpha$ with the $z$-axis defined by the uniform ambient magnetic
field (of magnitude $B_0$).
The total plasma density field is normalized by $\rho_0$, the
magnetic field by $B_0$, the  velocities by 
the Alfv\'en velocity $v_A=B_0/(4\pi\rho_0)^{1/2}$, the 
pressures by the parallel ion pressure $p_0=p_\|^{i (0)}$,  the heat
fluxes by $p_0 v_A$ and the fourth rank moments by $p_0
v_A^2$. The unit of length is the ion inertial length
$v_A/\Omega_i$, and time is normalized to ion gyroperiods. The parameter $\beta=8\pi p_0/B_0^2$ measures the ratio
of the (parallel) thermal to the magnetic pressure. Velocities without
superscripts refer to the ion velocity. The electron velocity $u^e$
is given by
\begin{eqnarray}
&&u_x^e=u_x+\cos\alpha\frac{\partial_\xi b_y}{\rho}\\
&&u_y^e=u_y-\frac{\partial_\xi b_p}{\rho}\\
&&u_z^e=u_z-\sin\alpha\frac{\partial_\xi b_y}{\rho} ,
\end{eqnarray}
where we define $b_p=\cos\alpha\ b_x-\sin\alpha\ b_z$.
We also define  $\bar u=\sin\alpha\ u_x+\cos\alpha\ u_z$ which, together
with  $\nabla\cdot u=\partial_\xi\bar u $, take the same form when using the electron
velocity. When integrated, the divergenceless condition $\nabla\cdot b=0$ rewrites
$\cos\alpha\ b_z+\sin\alpha\ b_x=\cos\alpha$, which allows
to write $\nabla\cdot\widehat
b=\nabla\cdot(b/|b|)=-\cos\alpha\partial_\xi |b|/|b|^2$.

The model involves dynamical equations for the ion
density $\rho$ and velocity $u$, the magnetic field components
$b_p$ and $b_y$, together with, for each species $r$, the gyrotropic
parallel and perpendicular pressures 
$p_\|^r$ and $p_\perp^r$, the heat fluxes $q_\|^r$ and
$q_\perp^r$ and the mixed fourth order cumulants $\tilde r_{\|\perp}^r$. They read
\begin{eqnarray}
&&\partial_t\rho+\partial_\xi (\rho\bar u)=0\\
&&\partial_t (\rho u_j)+\partial_\xi (F_j^1+F_j^2)=0\\
&&\partial_t b_p=\partial_\xi E_y\\
&&\partial_t b_y=-\partial_\xi E_p\\
&&\partial_t p_\|^r=-\partial_\xi (\bar u
  p_\|^r+\frac{\cos\alpha}{|b|}q_\|^r+\sin\alpha S_x^{\| r}
  )\nonumber\\
&&\quad\qquad  +2q_\perp^r\nabla\cdot\widehat b-2p_\|^r\widehat b\cdot\nabla
  u^r\cdot\widehat b\\
&&\partial_t p_\perp^r=-\partial_\xi (\bar up_\perp^r
+\frac{\cos\alpha}{|b|}q_\perp^r+\sin\alpha S_x^{\perp r}
)\nonumber\\
&&\quad\qquad -q_\perp^r\nabla\cdot\widehat b-p_\perp^r\nabla\cdot
u+p_\perp^r\widehat b\cdot\nabla 
  u^r\cdot\widehat b\\
&&\partial_t q_\|^r=-\partial_\xi (\bar u  q_\|^r+
\frac{\cos\alpha}{|b|} \tilde  r_{\|\|}^r)-\frac{3\beta}{2}p_\|^r\frac{\cos\alpha}{|b|}\partial_\xi
  (\frac{p_\|^r}{\rho^r}) \nonumber\\
&&\quad\qquad -3q_\|^r\widehat b\cdot\nabla
  u^r\cdot\widehat b+3\tilde r_{\|\perp}^r\nabla\cdot\widehat b\\
&&\partial_t q_\perp^r=-\partial_\xi (\bar u  q_\perp^r+
  \frac{\cos\alpha}{|b|} \tilde
  r_{\|\perp}^r)- \frac{\beta}{2}p_\|^r\frac{\cos\alpha}{|b|}\partial_\xi
  (\frac{p_\perp^r}{\rho^r})\nonumber\\
&&\quad\qquad -q_\perp^r\nabla\cdot u
  -(\frac{\beta p_\perp^r}{2\rho^r}(p_\|^r-p_\perp^r)+\tilde
  r_{\|\perp}^r-\tilde  r_{\perp\perp}^r) \nabla  \cdot\widehat
  b\nonumber\\
&&\quad\qquad - R_{NG}^r \\
&&(\partial_t+\bar u\partial_\xi)\tilde r_{\|\perp}^i={\cal F}^{-1}\Big
    (-{\overline T}_\|^i
    {\cal R}_{\|\perp 2}^i ik\widehat q_\perp^i
-\sqrt{{\overline T}_\|^i} {\cal R}_{\|\perp 1}^i |k| \widehat
    {\tilde r}_{\|\perp}^i   \nonumber\\
&&\qquad +{\overline T}_\|^i{\overline T}_\perp^i
    ({\overline T}_\perp^i-{\overline T}_\|^i)\cos\alpha\sin\alpha{\cal R}_{\|\perp 3}^i \
    k^2({\cal C}_1+1)   \widehat b_y \Big )\label{eq:rparperpp}\\
&&(\partial_t+\bar u\partial_\xi)\tilde r_{\|\perp}^e={\cal F}^{-1}\Big
    (-{\overline T}_\|^e
    {\cal R}_{\|\perp 2}^e ik\widehat q_\perp^e
-\sqrt{{\overline T}_\|^e} {\cal R}_{\|\perp 1}^e |k|\widehat
    {\tilde r}_{\|\perp}^e   \nonumber\\
&&\qquad -{\overline T}_\|^e{\overline T}_\perp^e
    ({\overline T}_\perp^e-{\overline T}_\|^e)\cos\alpha\sin\alpha{\cal R}_{\|\perp 3}^e \
    k^2 \widehat b_y \Big ).\label{eq:rparperpe}
\end{eqnarray}
The electric field and flux terms entering the above equations are
given by
\begin{eqnarray}
&&E_p=\cos\alpha\ E_x-\sin\alpha\ E_z\\
&&E_j=-(u\times b)_j -\frac{1}{\rho}\partial_\xi F_j^2\\
&&F_x^1=\rho\bar u u_x+\frac{\beta}{2}\Big (\sin\alpha\
  p_\perp^i+\cos\alpha\
  (p_\|^i -p_\perp^i)\frac{b_x}{|b|^2}\nonumber\\
&&\quad\qquad +\sin\alpha\ \Pi_{xx}+\cos\alpha\  \Pi_{xz}\Big )\\ 
&&F_y^1=\rho\bar u u_y+\frac{\beta}{2}\Big (\cos\alpha\
  (p_\|^i -p_\perp^i)\frac{b_y}{|b|^2}\nonumber\\
&&\quad\qquad +\sin\alpha\ \Pi_{xy}+\cos\alpha\  \Pi_{yz}\Big )\\ 
&&F_z^1=\rho\bar u u_z+\frac{\beta}{2}\Big (\cos\alpha\
  p_\perp^i+\cos\alpha\
  (p_\|^i -p_\perp^i)\frac{b_z}{|b|^2}\nonumber\\
&&\quad\qquad +\sin\alpha\  \Pi_{xz}\Big )\\ 
&&F_x^2=\sin\alpha\  \frac{|b|^2}{2}-\cos\alpha\
  b_x+\frac{\beta}{2}\Big (\sin\alpha\
  p_\perp^e\nonumber\\
&&\quad\qquad +\cos\alpha\ (p_\|^e -p_\perp^e)\frac{b_x}{|b|^2}\Big )\\
&&F_y^2=-\cos\alpha\  b_y+\frac{\beta}{2}\Big (
\cos\alpha\ (p_\|^e -p_\perp^e)\frac{b_y}{|b|^2}\Big )\\
&&F_z^2=\cos\alpha\  \frac{|b|^2}{2}-\cos\alpha\
  b_z+\frac{\beta}{2}\Big (\cos\alpha\
  p_\perp^e\nonumber\\
&&\quad\qquad +\cos\alpha\ (p_\|^e -p_\perp^e)\frac{b_z}{|b|^2}\Big ).
\end{eqnarray}
The operator ${\cal F}^{-1}$ is the inverse Fourier transform. The second terms
on the right-hand-side of Eqs. (\ref{eq:rparperpp}) and (\ref{eq:rparperpe}) involve a Hilbert transform with respect to the $z$-variable (which in
Fourier space reads ${\cal H} \hat f = i k/|k| \hat f$), signature of the Landau damping.
The specific form of the expressions for the gyrotropic fourth rank cumulants $\tilde
r_{\|\|}^r$ and $\tilde r_{\perp\perp}^r$, of  the gyroviscous tensor $\Pi$,
of the non-gyrotropic contributions to the fourth-rank cumulant 
$R_{NG}^r$, of the transverse components 
of the fluxes of parallel and perpendicular heat $S_x^{\| r}$ and
$S_x^{\perp r}$ as well as of the coefficients ${\cal R}_{\|\perp p}^r$ and
function ${\cal C}_1$,  are computed from the linear kinetic theory and given
in \cite{BPS07}. They involve functions $\Gamma_0$ and
$\Gamma_1$ where $\Gamma_\nu(b)$ is the product of $\exp(-b)$
by the modified Bessel function $I_\nu(b)$, with 
$b\equiv k_\perp^2 r_L^2/2=\displaystyle{\frac{\beta}{2}{\overline T}_\perp^i k^2 \sin^2\alpha}$.

%
%
%
%
%
%
%
%



%
%
%

\newpage

\end{document}